\documentclass{aa}

\usepackage{graphicx}
\usepackage{txfonts}
\usepackage{amstext}
\usepackage{upgreek}
\usepackage[mathlines,switch]{lineno}

\usepackage{cancel}
\usepackage[dvipsnames]{xcolor}

\newcommand{\bqn}{\begin{eqnarray}}
\newcommand{\eqn}{\end{eqnarray}}

\def\PGPU{$\varphi-$GPU }

\renewcommand{\arraystretch}{1.8}

\def\gapprox{\;\rlap{\lower 3.0pt                       
        \hbox{$\sim$}}\raise 2.5pt\hbox{$>$}\;}
\def\lapprox{\;\rlap{\lower 3.1pt                       
        \hbox{$\sim$}}\raise 2.7pt\hbox{$<$}\;}

\newcommand{\be}{ \begin{equation} }
\newcommand{\ee}{\end{equation}}

\newcommand{\ben}{\begin{enumerate}}
\newcommand{\een}{\end{enumerate}}

\makeatletter
\renewcommand*\aa@pageof{, page \thepage{} of \pageref*{LastPage}}
\makeatother
\newcommand{\orcid}[1]{\href{https://orcid.org/#1}{\protect\includegraphics[width=8pt]{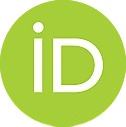}}}
\usepackage{hyperref}
\usepackage{ulem}
\hypersetup{colorlinks = true, citecolor = {blue}, urlcolor= {blue}, linkcolor={red}}

\linenumbers
\begin{document}

\nolinenumbers      

\title{Dynamical evolution timescales for the supermassive black hole system in the galaxy NGC~7727 ( Arp~222 )}

\titlerunning{Triple SMBH system in NGC~7727}
\author{
P.~Berczik\orcid{0000-0003-4176-152X}\inst{1,2,3}
\and
M.~Ishchenko\orcid{0000-0002-6961-8170}\inst{1,2,3}
\and
O.~Veles
\inst{1,2}\orcid{0000-0001-5221-2513}
\and
M.~Sobolenko\orcid{0000-0003-0553-7301}\inst{1,2}
\and
K. Voggel\orcid{0000-0001-6215-0950}\inst{4}
\and
C.M. Boily\orcid{0000-0002-7274-6720}\inst{4}
\and
E.~Polyachenko\orcid{0000-0002-4596-1222}\inst{5}
\and
R.~State\orcid{0000-0002-4751-9577}\inst{5} 
}

\institute{
Main Astronomical Observatory, National Academy of Sciences of Ukraine,
27 Akademika Zabolotnoho St, 03143 Kyiv, Ukraine, 
\email{\href{mailto:berczik@mao.kiev.ua}{berczik@mao.kiev.ua}}
\and
Nicolaus Copernicus Astronomical Centre Polish Academy of Sciences, ul. Bartycka 18, 00-716 Warsaw, Poland
\and
Fesenkov Astrophysical Institute, Observatory 23, 050020 Almaty, Kazakhstan
\and
Observatoire astronomique de Strasbourg and CNRS, UMR 7550, 11 rue de l'Université, Strasbourg 67000, France
\and
SnT SEDAN, University of Luxembourg, 29 boulevard JF Kennedy, 1855 Luxembourg, Luxembourg
}
   
\date{Received xxx / Accepted xxx }

\abstract    
{A dual active galactic nucleus candidate with a separation of only $\approx500$~pc was recently found in NGC~7727. According to the hierarchical merging scenario, such objects would be expected to merge on a timescale of a few hundred million years (Myr). However, estimating the accurate merging timescales for the two nuclei is still a complex challenge. }
{Using our numerical \textit{N}-body code, we can trace the full evolution of central black holes during all phases: dynamical friction of unbound black holes, binary black hole formation, hardening of the system due to two-body scattering, and emission of gravitational waves leading to the final merger.
}
{According to the extracted observational data, the numerical model has three main components: the bulge contains two dense stellar nuclei, each of which hosts a black hole. The observed system is in an advanced stage of merging, where the most massive black hole in the center of the galaxy has a mass of $1.54\times10^{8}\rm\;M_\odot$ and the least massive black hole in the offset second stripped nucleus has a mass of $6.33\times10^{6}\rm\;M_\odot$. Using a direct $N$-body \PGPU code, we followed the dynamical evolution of the system up to a final separation of four Schwarzschild radii. The black holes were added as special relativistic particles and their equation of motion contains a full post-Newtonian approximation up to the 2.5 term.}
{Initially, the black holes are not gravitationally bound and, thus, the system spends more than 60~Myr in the phase of dynamical friction while tightening the orbit. Then, a bound binary forms quickly with a relatively high eccentricity of $e\approx0.98$. The two-body scattering phase takes place from $\approx60$~Myr up to $\approx120$~Myr. In the last $\approx10$~Myr, the black hole's separation is seen to be rapidly shrinking due to the gravitational wave emission. Starting from the physical separation observed today, the total merging time in our model is $130\pm10$~Myr, which is less than half the value of the previous estimates.
}
{Our study points to the possibility of the binary black holes  of NGC~7727 merging on a relatively short timescale. These results have implications for the statistics of strong sources of gravitational waves at low frequencies, namely, systems engaged in an advanced state of merging (similarly to the case of NGC~7727) are expected to be  prime sources for the LISA mission to observe.}

     
\titlerunning{SMBH system in NGC~7727}
\authorrunning{P.~Berczik et al.}
\maketitle

\section{Introduction}\label{sec:Intr}

Nearly all massive galaxies contain a super-massive black hole (SMBH) in their centre, as detailed in the review by \cite{Kormendy2013}. However, the growth channels of SMBHs and their initial seed masses are not well understood. Black holes can grow through merging processes between two SMBHs or through gas accretion when they are in an active galactic nucleus  (AGN) state. In recent years, JWST has shed new light on SMBHs in the early universe, for instance, with the discovery of an active SMBH at a redshift of 10 \citep{Maiolino2024}; hence, the early stages of SMBH mass growth become more accessible to observations. From a theoretical modelling perspective, the merging of the SMBH in the galactic centres would be the bright source of gravitational wave emission \citep{Reisswig2009, Holley-Bockelmann2020, Volonteri2003}.

Despite the scepticism of the early 2000s about the SMBH binary (SMBHB) merging timescales (the so-called final-parsec problem), up-to-date realistic high-resolution cosmological and zoomed cosmological galactic-scale simulations point to a rather short timescale for the BHs merging in galactic centres. As leading examples of such works, we can refer to Illustris TNG50 \citep{Nelson2019} and Romulus25 \citep{Tremmel2017}. {In addition to the obvious great achievements of these simulations in the explanation of the formation of galactic-scale objects and long-term dynamical evolution, they also display some evident resolution issues (gravitational softening) with respect to  detailed investigations of the possible SMBH mergers (see details in \citealt{Khan2016}).} Also, based on cosmologically motivated zoomed high resolution simulations \citep{Khan2016, Khan2018ApJ, Koehn2023} confirmed the fast merging of the galaxies' nuclei. In this process, the central galactic BHs are transferred to the common centre of the forming system on a dynamical friction timescale. 

The SMBH binary merging process in galactic centers can dynamically be separated into a few stages \citep{Begelman1980}. During the initial stage of the galaxy merger, the SMBHs fall to the centre of the system due to dynamical friction, which can be expressed as  

\begin{equation}
\dot{V}_{\rm df} = \frac{-4\pi G^2\rho M_{\rm bh}}{V_{\rm bh}^2}\chi 
        \ln\Lambda
    \quad \mathrm{with}\quad 
    \chi=\frac{\rho(<V_{\rm bh})}{\rho}.
    \label{dynfric}
\end{equation}

In general, the functions, $\chi$, and the Coulomb logarithm, $\Lambda$, depend on the velocity of the massive object and the properties of the background system \cite{BT1987}. For more details on this initial dynamical friction-dominated stage, we refer to \cite{Just2011}. As a final product of this stage, we obtained the system of initially loosely bound BHs embedded in a large number of field stars.

The next dynamical stage is the tightening of the SMBH binary through repeated energy exchanges with the background stars \citep{Merritt2001a, Merritt2001b}. At the end of this stage, we expect to see the hard binary SMBH already close to the mpc separation where the relativistic post-Newtonian effects have the capacity to further shrink the binary up to the GW merging phase \citep{Brem2013, Sobolenko2017, Gualandris2022, Khan2015}. In addition, more details on the available regularisation techniques to follow dynamical friction, scouring, and post-Newtonian evolution without gravitational softening in galaxy-scale and cosmological simulations can be found in literature \citep{Rantala2017, Mannerkoski2023, Partmann2024, Zhou2025}. 

\cite{Tamfal2018}, \cite{Liao2024}, \cite{Gualandris2022}, and \cite{Rawlings2023} explored the role of initial conditions and environmental factors in determining density profiles, nuclear star formation, and stochastic effects. During the final merging phase of the galactic SMBH, stochastic or random effects such as gravitational fluctuations are present. Such effects typically lead to the formation of a hard binary characterised by high eccentricity, whose evolution continues to become even more eccentric (for more details see  Figs. 5 in \citealt{Gualandris2022} and \citealt{Sobolenko2022}). A substantial body of numerical investigations has also been devoted to the interaction of the forming SMBH binary embedded in a mostly gaseous circum-nuclear or circum-binary disk \citep{Fiacconi2013, Mayer2013, Farris2014, Ryan2017}.

NGC~7727 (Arp~222) is a peculiar spiral galaxy located about 27~Mpc away in the constellation Aquarius  \citep{Schweizer2018}. It is notable for its strong tidal features, which are likely the result of a recent merger. Recently, it was discovered that the galaxy contains two separated nuclei and each of them hosts a SMBH \citep{Voggel2022}. One nucleus (hereafter {\tt Nuc~1}) is in the photometric centre of the galaxy, whereas the second nucleus ({\tt Nuc~2} hereafter) is offset from the centre by only 500~pc. The second nucleus is probably the stripped nuclear star cluster of the merged galaxy. In \citet{Voggel2022}, it was predicted that due to the small apparent separation of the two nuclei, they will likely merge within the next 250~million years. Therefore,  this system could offer a rare, detailed insight into a double SMBH system that is close to merging its two SMBHs under the emission of gravitational waves \citep{Moore2015}.

The galaxy's proximity to Earth  makes it a great study case for galaxy merging and interaction as well as for the merger of two SMBHs. With the upcoming LISA mission \citep{Moore2015}, the understanding of double SMBH systems has become even more important, as their frequency is a direct information needed in the prediction of LISA event rates.

At the time of discovery, this was the closest known double SMBH system and, even more crucially, the closest in distance to us.  This proximity enables us to study it in unprecedented detail, revealing a wealth of dynamical information that is not accessible with other, more distant candidates (e.g. \citealt{Kollatschny2020}).

The goal of this paper is to determine the shortest time required for the two SMBHs to merge. Earlier estimates based on a classic Chandrasekhar dynamical friction set the time of orbital migration by either of the SMBHs to be of the order of 250~Myr, for more details see Section 6.1 in \cite{Voggel2022}. These estimates do not take into account the impact of anisotropic stellar velocities surrounding the SMBHs, which could be due, for example,  to the residual rotation of the galactic system or streams of stars still observed on a kpc scale. The physical distance of $\approx 500$~pc seen in the projection does not set strong constraints on the current orbit of the two sub-systems. In this contribution, we hypothesise that the projected distance is close to the actual three-dimensional relative positions, so that the orbit is in the plane of the sky. 

{The paper is organised as follows. In Section~\ref{sec:model}, we briefly describe the physical characteristics of NGC~7727, construct a simple physical model and corresponding numerical models. We also briefly discuss an $N$-body numerical code for the simulations. In Section~\ref{sec:dyn-evol} we present the dynamical timescales for the modelled SMBHB and depict their GW signal. We also discuss merger statistic from such types of objects. In Section~\ref{sec:con}, we summarise our results and conclusions.}

\section{Model construction for NGC 7727}\label{sec:model}

\subsection{Physical properties of NGC 7727}

\begin{figure}
\centering
\includegraphics[width=0.98\linewidth]{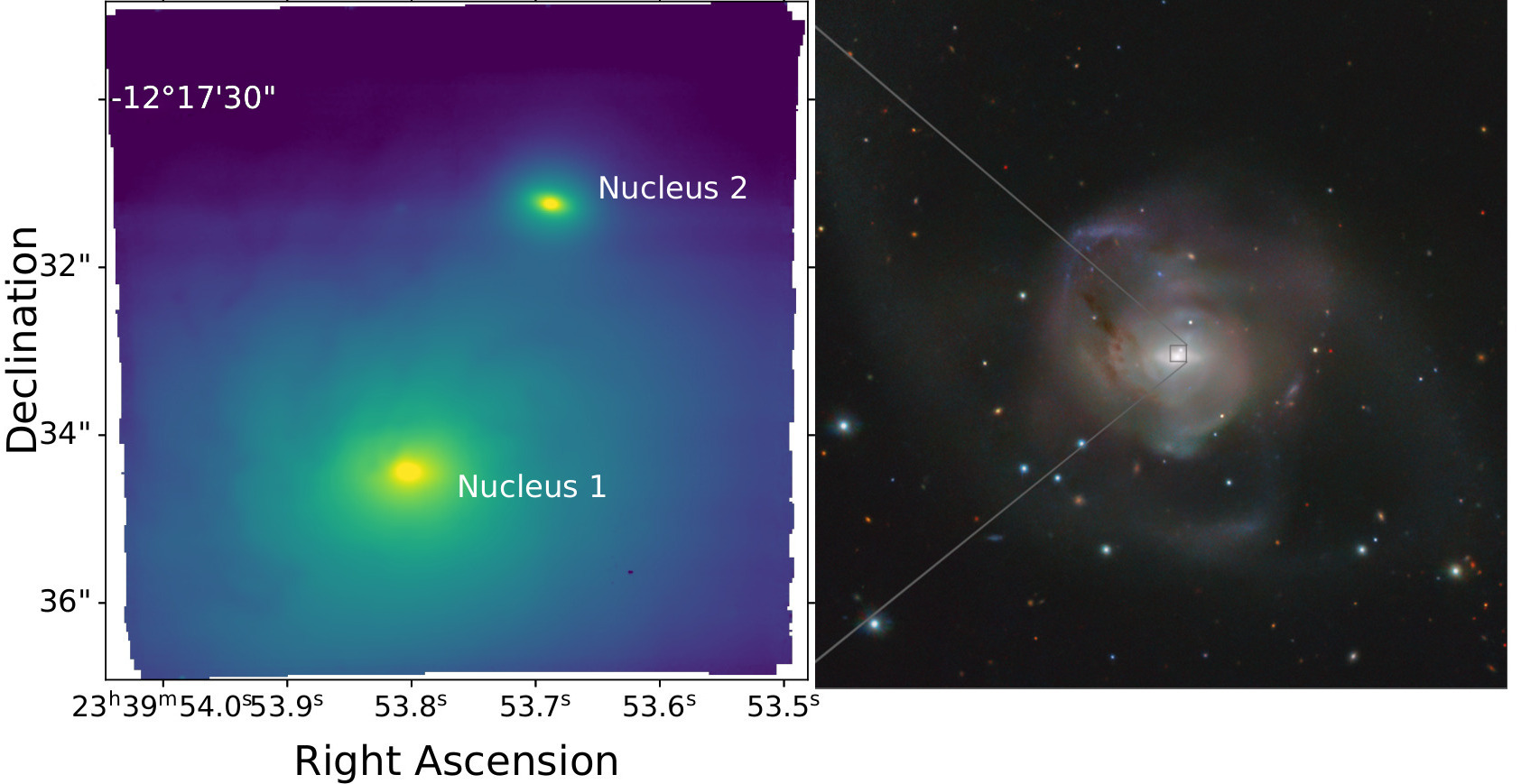}
\caption{Visualisation of the NGC~7727. Left image is  from \cite{Voggel2022}. Right image is  from ESO's VLT Survey Telescope.} 
\label{ngc-7727-img}
\end{figure}

In a recent and comprehensive study of NGC~7727, \cite{Voggel2022} combined VLT kinematic data with HST photometry and Jeans anisotropic equilibrium galaxy models to confirm the presence of an SMBH in both nuclei.  According to observational data, NGC~7727 has a total stellar mass of $M_{\ast}=1.3\times10^{11}\rm\;M_{\odot}$ and resides in a dark matter (DM) halo with a mass of around $10^{13}\rm\;M_{\odot}$ and a typical  stellar-to-halo-mass relation \citep{Behroozi2013}. 

\cite{Voggel2022} also determined the masses of the SMBHs using high spatial resolution spectroscopy from the MUSE instrument. They found that {\tt Nuc~1}, located at the photometric centre of the galaxy, hosts a black hole with a mass of $M_1=1.54\times10^{8}\rm\;M_{\odot}$. The second nucleus {\tt Nuc~2} contains a smaller black hole with mass $M_2=6.33\times10^{6}\rm\;M_{\odot}$. A careful assessment of the background contamination by field stars led  \citet{Voggel2022} to rule out a BH-free {\tt Nuc~2} at a $4.5\sigma$~confidence level. The two SMBHs are characterised by a mass ratio of $\approx 1:24$, which makes the NGC~7727 a typical example of the minor merger SMBH system. 

With an integrated bulge mass of $2.1 \times 10^8\rm\;M_{\odot}$, the {\tt Nuc~2}'s SMBH makes up to 4.5\% of the second nucleus, well above  the $\sim 0.1 \% $ expected from the BH-bulge scaling relation \citep{Kormendy2013}. 
For that reason, {\tt Nuc~2} has been identified as the surviving nuclear star cluster left over from a smaller galaxy that merged with NGC~7727. Additionally, optical emission lines indicate that {\tt Nuc~2} is a Seyfert-type low-luminosity AGN. The spread in colours and age templates reaching down to $\simeq1.5$~Gyr makes it likely that the progenitor galaxy carried with it a gas-rich disc, which accounts for the younger stellar population forming throughout the merger. Thus {\tt Nuc~2} is in an advanced in-spiral phase, which will eventually lead to the merger of the two SMBHs: analytical estimates relying on Chandrasekhar dynamical friction suggest that the final binding and merger of the two nuclei will take place in less than 1~Gyr, based on  a plausibly short 250~Myr timescale. 

\subsection{NGC 7727 Initial condition}\label{subsec:phys-mod}

To build the initial physical model, the estimated parameters of both nuclei and SMBHs, along with the stellar and gas components, were collected from existing observational data \citep{Schweizer2018, Voggel2022}. 

The stellar mass of {\tt Nuc~1} is  $1.14\times 10^9\rm \; M_\odot$, while the {\tt Bulge} is $51.6\times 10^9 \rm \;M_\odot$, giving us a total mass of around $52.7\times 10^9 \rm\; M_\odot$. We estimated the mass of  {\tt Nuc~2} as $0.2\times 10^9 \rm \;M_\odot$. According to \cite{Voggel2022}, the observed masses of the SMBHs are $M_{\rm BH1} = 1.54\times10^{8}\rm \; M_\odot$ and $M_{\rm BH2} = 6.33\times10^{6} \rm\;M_\odot$. The  fitted physical parameters of the central part of NGC~7727 are summarised in Table~\ref{tab:init-par7727}.  The observed surface brightness distribution of the combined three S\'{e}rsic profiles is given in Fig.~\ref{ngc-7727-img}. 

We initiated the galaxy merger from the dynamical system of two unbound central SMBHs located inside {\tt Nuc~1} and {\tt Nuc~2} with an initial separation of $\Delta R = 480$~pc. As the type of orbit for the SMBH2, we chose an eccentric one (assuming that we are now at the pericentre) with orbital eccentricity {\tt e} = 0.5.
%
{To test the assumptions made about the physical configuration of the two nuclei's positions and orbital eccentricity, along with their impact on the outcome of our models, we performed calculations (not included in this contribution) with a wider initial physical separation between the nuclei obtained for a projection angle $i$ varying from 30 to 60~degrees, increasing the 
separation from 554~pc up to 960~pc. The main effect is to 
delay the formation of the binary by up to at most 35~Myr. 
Hence, our choice of $i = 0^{\rm o}$ clearly leads to the shortest merger timescale and provides a limiting case at the same time.}

\begin{table*}
\caption{NGC~7727 physical model parameters.}
\label{tab:init-par7727}
\centering
\renewcommand{\arraystretch}{1.2}
\begin{tabular}{cccccccc}
\hline
\hline
Component & $M_{*}$ & $M_{\bullet}$ & Profile & Effective  &  S\'{e}rsic & Axis & Axis \\
 & $10^{9}, \rm\;M_{\odot}$ & $10^{7}, \rm\;M_{\odot}$ & & radius, pc & index & ratio, Y &  ratio, Z \\
\hline
\hline
{\tt Nuc~1} & 1.14 & 15.4 & S\'{e}rsic & 47.1 & 1.51 & 1.0 & 0.6 \\
{\tt Nuc~2} & 0.2 & 0.633 &  S\'{e}rsic & 30.1 & 1.79 &  1.0   & 0.62 \\
{\tt Bulge}  & 51.6 & -- &  S\'{e}rsic & 811.0 & 2.28 & 1.0 & 0.74 \\
\hline
\end{tabular}
\begin{minipage}{\linewidth}
\smallskip
\end{minipage}
\end{table*}

\subsection{Numerical realizations of the NGC~7727 initial condition}\label{sybsec:num-real}

In our physical model, each of the SMBHs is surrounded by its own initially bound stellar systems {\tt Nuc~1} and {\tt Nuc~2}, which are located in the common stellar system known as the {\tt Bulge}. All three density distributions are described by the generalised S\'{e}rsic model \citep{Sersic1963}. For the numerical realisation of the S\'{e}rsic distribution of each of the three components, we used the popular {\tt Python}-based  {\tt AGAMA} library \citep{agama2019}. The main parameters of the distributions we present in Table~\ref{tab:init-par7727}. 
We centred the {\tt Nuc~1} and {\tt Bulge} systems in the origin of the coordinate system. We placed the {\tt Nuc~2} system at an initial distance of $\Delta R = 480$~pc along the $Z$ axis. We note that we generated the combined system with three different S\'{e}rsic index models and  with three slightly different axis flattening modes as well. 

\begin{table}
\caption{NGC~7727 numerical models.}
\label{tab:num-mod7727}
\centering
\renewcommand{\arraystretch}{1.2}
\begin{tabular}{cccc}
\hline
\hline
Component & \multicolumn{3}{c}{Particles number, $N$} \\
 &  Model {\tt A} & Model {\tt B} & Model {\tt C} \\
\hline
\hline
{\tt Nuc~1} & 57\textit{k} & 114\textit{k} & 228\textit{k} \\
{\tt Nuc~2} & 10\textit{k} & 20\textit{k} & 40\textit{k} \\
{\tt Bulge} & 258\textit{k} & 516\textit{k} & 1 032\textit{k} \\
\hline
Total & 325\textit{k} & 650\textit{k} & 1 300\textit{k} \\
\hline
Rand. seed & rnd1, rnd2  & rnd1, rnd2  & rnd1, rnd2 \\
\hline
\end{tabular}
\end{table}

To check the robustness of the numerical modelling results for the dynamical SMBHB orbital hardening, we constructed three different numerical realisations of the same physical model with three different particle numbers: $N$ = 325\textit{k}, 650\textit{k}, and 1300\textit{k} (see Table~\ref{tab:num-mod7727} with the numerical indexes {\tt A}, {\tt B}, {\tt C)}. The particle numbers for each component were chosen in such a way that the individual masses of field particles would end up with nearly the same masses. In this way, we were able to suppress the mass segregation effects (due to the different masses of particles from different components) in our dynamical N-body system during evolution. 

\begin{figure}
\centering
\includegraphics[width=0.99\linewidth]{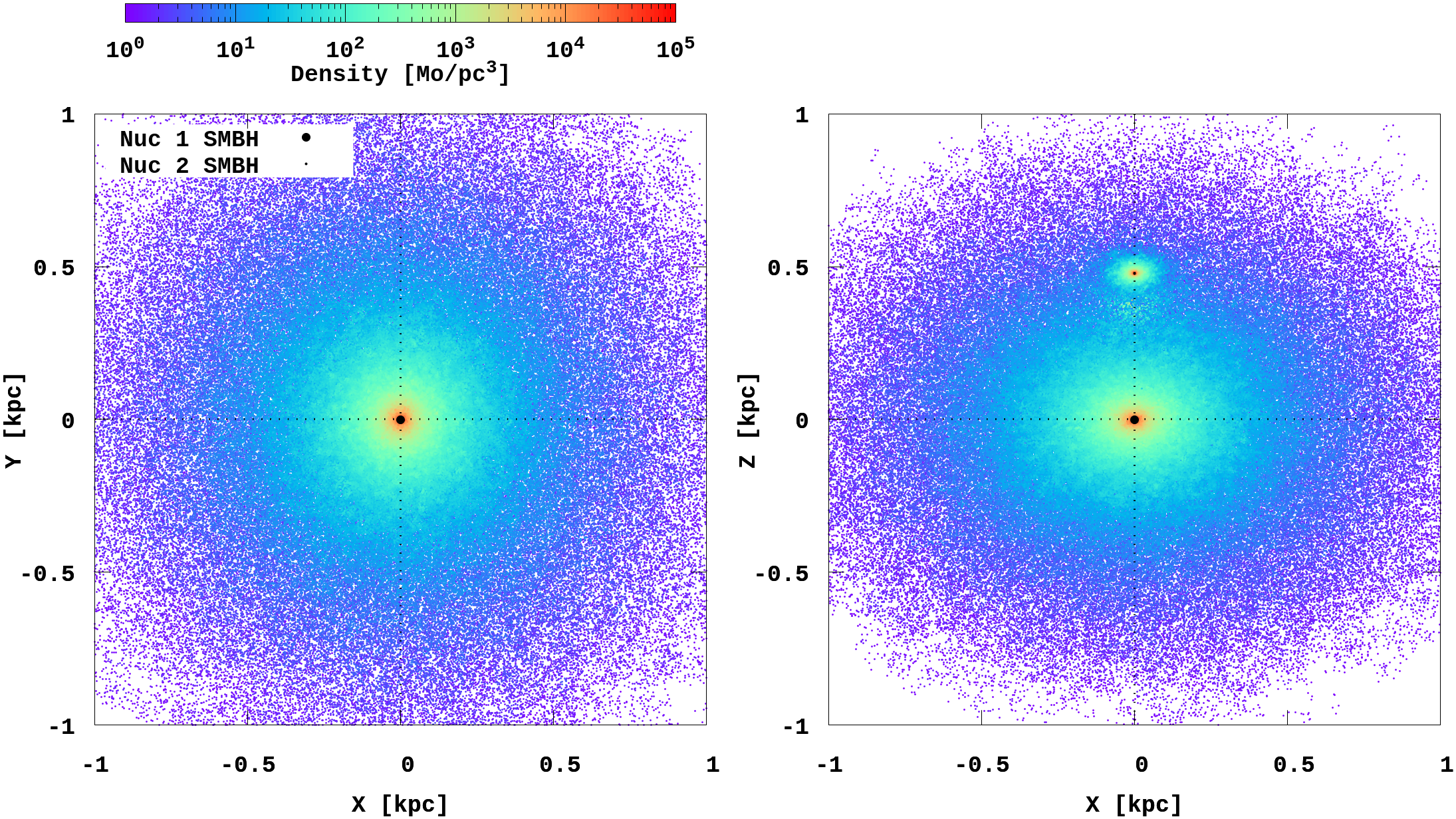}
\caption{Initial density distributions at projection planes ($X$, $Y$) and ($Y$, $Z$) for numerical model {\tt B} = 650\textit{k}, rnd2. Black and dark blue dots represent the BHs in {\tt Nuc~1} and {\tt Nuc~2}, respectively.}
\label{fig:pos-den}
\end{figure}

For each of these numerical realisations, we created two particle distributions with two randomised initial positions and velocities. Thus, in total, we obtained six independent numerical realisations of the same physical model, as detailed in Table~\ref{tab:num-mod7727}. In Fig.~\ref{fig:pos-den}, we present the density distribution at the initial time for our physical NGC~7727 configuration for the numerical realization {\tt B} = 650\textit{k}, rnd2.

\begin{figure}
\centering
\includegraphics[width=0.99\linewidth]{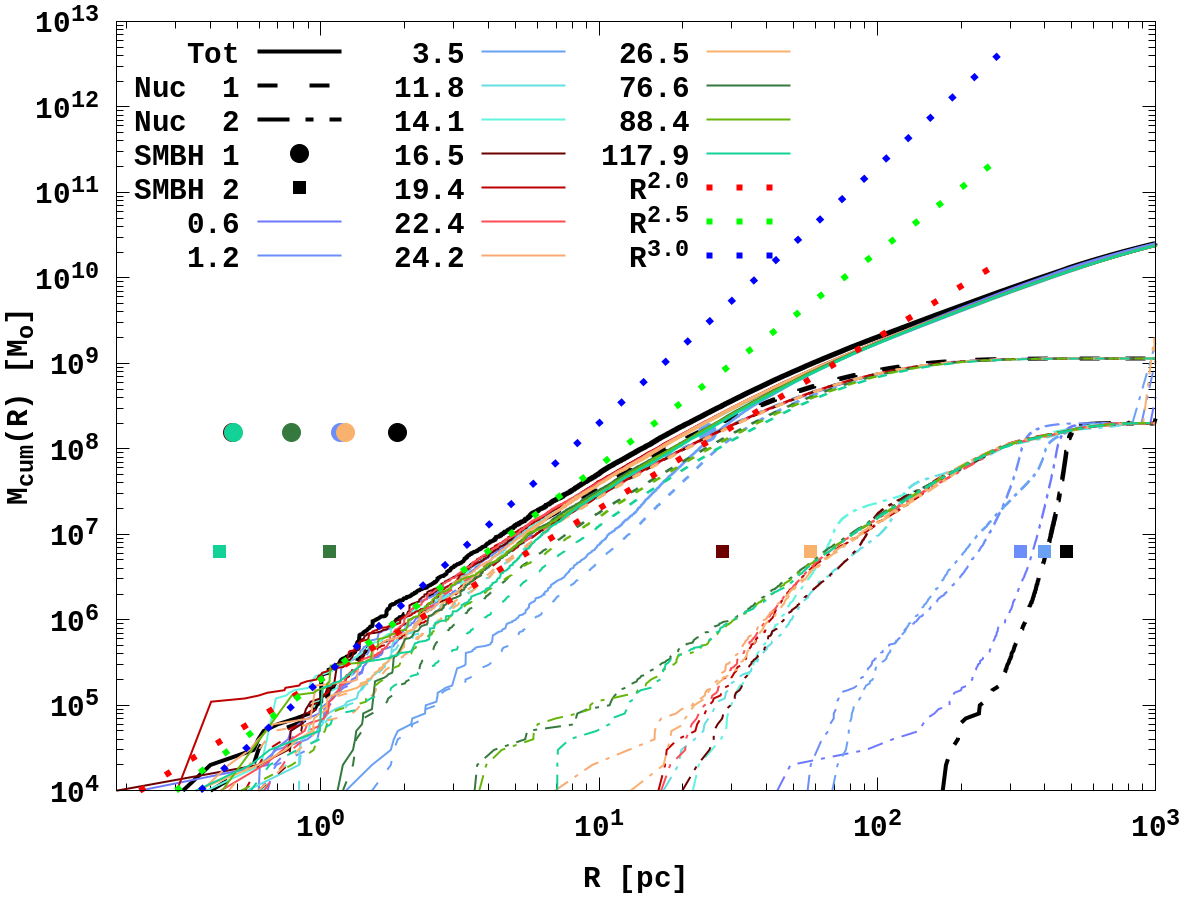}
\caption{Evolution of the total cumulative mass distribution  (black lines for the initial masses), for the bound stage (blue shaded lines), for the merged stage (red shaded lines), and post-Newtonian phase (green shaded lines) for model {\tt B$_{\tt rnd2}$}. The labels on the shades besides black represent the passed time in the simulation in Myrs.}
\label{fig:cum-m}
\end{figure}

In Fig.~\ref{fig:cum-m}, we present the evolution of the cumulative mass distribution profiles for the three different components {\tt Nuc~1}, {\tt Nuc~2}, and {\tt Tot} = {\tt Nuc~1 + Nuc~2 + Bulge}. The cumulative mass distribution at the start of the simulation is presented as black solid and dashed lines. The two SMBHs' individual positions and masses are also represented at the same time. 

During the $N$-body modelling, we used the NB scaling units (as with H\'{e}non units\footnote{H\'{e}non units: \\~\url{https://en.wikipedia.org/wiki/N-body_units}}, here we set the gravitational constant G equal to 1). These numerical simulation units are based on the main observed physical parameters of the system; that is, the distance between the components {\tt Nuc~1} and {\tt Nuc~2} and the total mass of the system. In this way, we obtained the next NB scaling units for mass, distance, velocity, and time, respectively expressed as 

$M_{\rm NB} = 1.0\times 10^{{9}}\rm\;M_{\odot}$,\;\;\;\;\;\;\;\;\;\;\; 
$R_{\rm NB} = 100$~pc,

$V_{\rm NB} \;= 207.4$~km~s$^{-1}$,\;\;\;\;\;\;\;\;\;\;\;\;
$T_{\rm NB} = 0.471$~Myr.

\noindent In these units, the speed of light velocity is $c_{\rm NB}$ = 1445.6. 

\subsection{N-body code}\label{sybsec:num}

For the global dynamical integration, we used the high-order parallel $N$-body code \PGPU\footnote{$N$-body code \PGPU:\\~\url{ https://github.com/berczik/phi-GPU-mole}}, which is based on the fourth-order Hermite integration scheme with hierarchical individual block time steps \citep{Berczik2011, BSW2013}. One more feature in our integration scheme is the minimisation and forced synchronisation of the SMBH particle time steps. In this way, we could ensure that the SMBH particle times were synchronised with each other and this also set a minimum time step between all particles. We already thoroughly tested and used our modified \PGPU code for a wide range of models that involve the evolution of multiple SMBHs in our previous works \citep{Sedda2019, Koehn2023, Berczik2024}. 

In the current version of the \PGPU code, we used individual softening for different types of particles. For the mixed interaction between different types of particles, we used the equation for effective softening: $\varepsilon_{\tt ij} = \varepsilon_{\tt ji} = 0.5 \times (\varepsilon_{\tt i} + \varepsilon_{\tt j})$. For the {\tt Nuc~1 + Nuc~2} particles, we set the gravitational softening $\varepsilon_{\tt Nuc} = 0.01$~pc and for the {\tt Bulge} we set $\varepsilon_{\tt Bulge} = 0.1$~pc. As a result, for the mixed softening between {\tt Bulge} and {\tt Nuc} particles, we get $\varepsilon_{\tt Nuc,Bulge} = 0.055$~pc. For the SMBH interactions, we applied exactly zero softening for the SMBH particles:  $\varepsilon_{\rm BH}=0$~pc. We used the same mixed (effective) softening for any interaction between field particles (star) and SMBH gravitational interactions, but with a special extra reduction factor: $\varepsilon_{\tt BH, i} = 0.01 \times 0.5 \times (\varepsilon_{\tt BH} + \varepsilon_{\tt i})$. The effective softening for these types of interactions is given as $\varepsilon_{\tt BH,Nuc} = 5 \cdot 10^{-5}$~pc and $\varepsilon_{\tt BH,Bulge} = 5 \cdot 10^{-4}$~pc, respectively. Initially, we ran the simulation in a pure Newtonian regime and turned on the post-Newtonian (PN) terms during the progressive merging stage and stopped the simulation when the separation of the SMBHs fell below four Schwarzschild radii. This final time was assumed to be the merging time for the SMBHB. In the current implementation of the code, we added the PN formalism up to the 3.5PN term (including the spin-orbit and spin-spin interactions as well) for the relativistic orbit calculation of the SMBHs \citep{Sobolenko2017, Sobolenko2022}. Among the PN terms,  2.5PN is the leading energy dissipation term, which `carries away' most of the binary SMBHB energy from the system due to the emission of gravitational waves \citep{Khan2018, Khan2018ApJ, Berczik2024}. In our current set-up, we used the non-Spin version of the PN formalism and limited the PN calculation up to the 2.5PN term.

\section{Dynamical evolution and gravitational waves}\label{sec:dyn-evol}

\subsection{Bound stage}\label{subsec:bound}

\begin{figure*}
\centering
\includegraphics[width=0.99\linewidth]{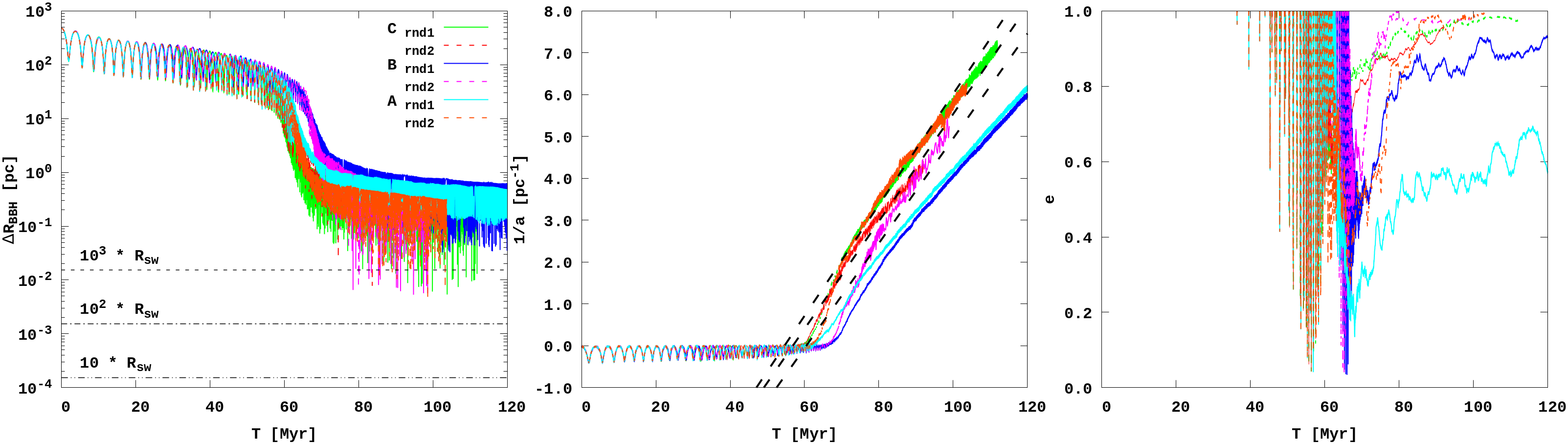}
\caption{Evolution of the black hole binary parameters at the stage of the pre-bound and bound systems. Left panel:\ Separations, $\Delta R$, between SMBHB for the numerical models {\tt A}, {\tt B,} and {\tt C} as solid lines. The dotted colour lines represent the second initial randomisation seeds for the particle distributions. Horizontal black dashed lines represent the level of the Schwarzschild radii for SMBHB with summarised mass: $M_{\rm \tt SMBHB}=M_{\rm \tt BH1}+M_{\rm \tt BH2}$. Middle pane: Evolution of the inverse semimajor axis for the same models. Black dashed lines show the simple linear fit for the `hardening' of the orbits for {\tt A}, {\tt B}, and {\tt C} models, respectively. Right panel:  Evolution of eccentricity, {\tt e}, for the SMBHB system for all numerical models.}
\label{fig:7072-bound}
\end{figure*}

We show the evolution of the SMBHB orbit parameters, such as the separation, $\Delta R$, inverse semi-major axis, $1/a$, and eccentricity, {\tt e}, in Fig.~\ref{fig:7072-bound} as part of our analysis of the evolution of the binary system. At $T = 0.0$~Myr, the SMBHs separated by $\Delta R$ are not gravitationally bound initially. 

In all of our models, the SMBHBs become bound after $\approx$60-65~Myr of dynamical evolution (Table~\ref{tab:boun-mer}), showing a quick inspiral time in the pure Newtonian regime (middle panel of Fig.~\ref{fig:7072-bound}). After this time, the inverse semi-major axis is constantly increasing almost as a linear function. The six numerical models presented here (in addition to significantly different particle numbers and different randomisation seeds for particle distributions) have consistent hardening rates $\approx0.15$~pc$^{-1}$~Myr$^{-1}$. We then see the formation of the hard binary within a timescale of $\approx 30$~Myr afterwards, with a high eccentricity of {\tt e} > 0.9, except in one model ({\tt A$_{\tt rnd1}$}) where it remained smaller. In our case, this high eccentricity also strongly influences the final PN merging time of the system \citep{Peters1963}. 

We conclude that all numerical models lead to a quite similar SMBHB evolution, independent of the variations in particle number or the range of random seed realizations. In all models, the pre-bound phase takes roughly the same time ($\approx60-65$~Myr). After this phase, the Newtonian hardening phase (due to the three-body scattering of stars around SMBHB) has a similar hardening rate (close to $\approx0.15$~pc$^{-1}$~Myr$^{-1}$). Therefore, after $\approx65-70$~Myr of evolution, we arrive at similar conditions for our PN simulations in the final merging stage.

\begin{table}
\caption{NGC~7727 bound and merger times.}
\label{tab:boun-mer}
\centering
\renewcommand{\arraystretch}{1.2}
\begin{tabular}{ccccccccc}
\hline
\hline
Models & N & $T_{\rm {bound}}$ & $T_{\rm {hard}}$ & $T_{\rm {merge}}$ \\
 & & Myr & Myr & Myr \\
\hline
\hline
{\tt A$_{\rm rnd1}$} & 325\textit{k} & 62 & 100 & 125 \\
{\tt A$_{\rm rnd2}$} & $\cdots$ & 64 & -- & -- \\
{\tt B$_{\rm rnd1}$} & 650\textit{k} & 66 & 103 & 119  \\
{\tt B$_{\rm rnd2}$} & $\cdots$ & 65 & -- & -- \\
{\tt C$_{\rm rnd1}$} & 1 300\textit{k} & 60 & 110 & 141 \\
{\tt C$_{\rm rnd2}$} & $\cdots$ & 63 & -- & -- \\
\hline
\end{tabular}
\end{table}

In Fig.~\ref{fig:cum-m}, we present the detailed evolution of the cumulative mass distribution of our system. The initial distribution is presented in black. The prebound epoch is presented in a dark blue colour. The distributions during the 
bound stages are plotted in light blue colour. In the hard-binary phase, we present the distributions as a red palette, while green shows the results for the final PN phase of our runs. During the complex dynamical evolution of our NGC~7727 model, we see a slight change in the slope of the combined total cumulative mass distribution {\tt Tot} from $\approx2$ to $\approx3$. This slope change can be interpreted as the formation of a dense core inside the central $\approx1 - 2$~pc of our simulation. 

\subsection{Merge stage}\label{sec:PN}

After the bound stage formed, we activated the PN description in the code (1PN, 2PN, and 2.5PN) for the three models {\tt A, B, C}, with the first set of randomization {\tt rnd1}. All three runs, with different particle numbers, produce the final merging of SMBHB around the same time. The PN terms were switched on between $\approx65-70$~ Myr, then the SMBHB orbital pericentre separation was smaller than a few thousand Schwarzschild radii. In all systems, the final merging occurs at $\approx120-140$~Myr (see also Table~\ref{tab:boun-mer}). 

\begin{figure*}
\centering
\includegraphics[width=0.90\linewidth]{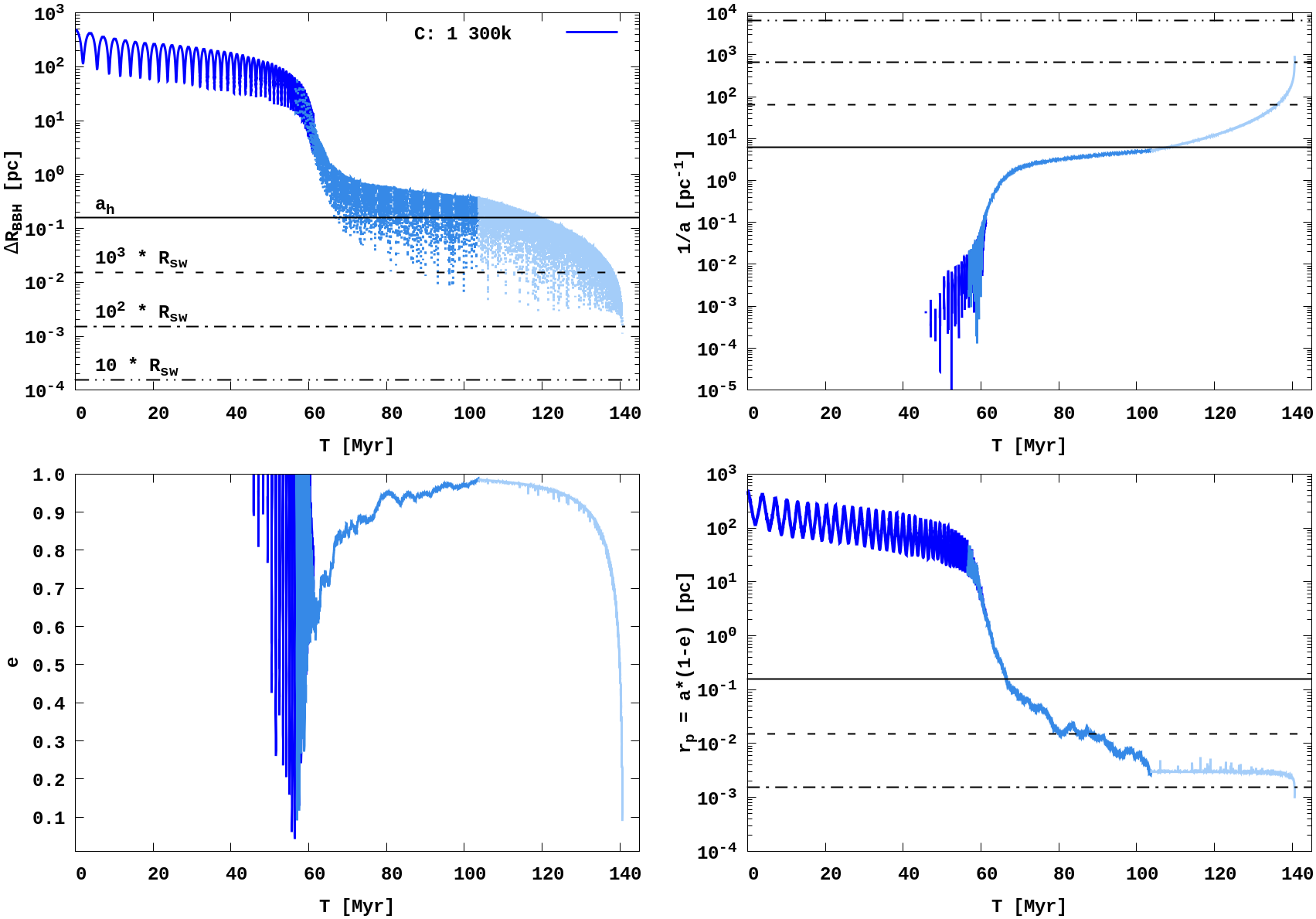}
\caption{Combined evolution of the black holes' parameters for the pre-merging and merging stages (including the PN phases) of the NGC~7727 SMBHB. Left upper panel: Separation $\Delta R$ between SMBHB for the numerical model {\tt C$_{\tt rnd1}$}. Horizontal black dashed lines represent 10, 100, and 1000 times the Schwarzschild radii for the SMBHBs' mass. The solid black line shows the size of the semi-major axis at the time when the binary starts to harden. Right upper panel: Evolution of the inverse semi-major axis for the same model. Left and right bottom panels: Evolution of eccentricity, {\tt e}, and pericentre for the SMBHB system. Different colours represent the different phases of the SMBHB evolution, with dark blue:\ unbound phase, blue:\ pre-merging, and light blue:\ merging, including  the PN terms.}
\label{fig:bh-param}
\end{figure*}

\begin{figure*}
\centering
\includegraphics[width=0.90\linewidth]{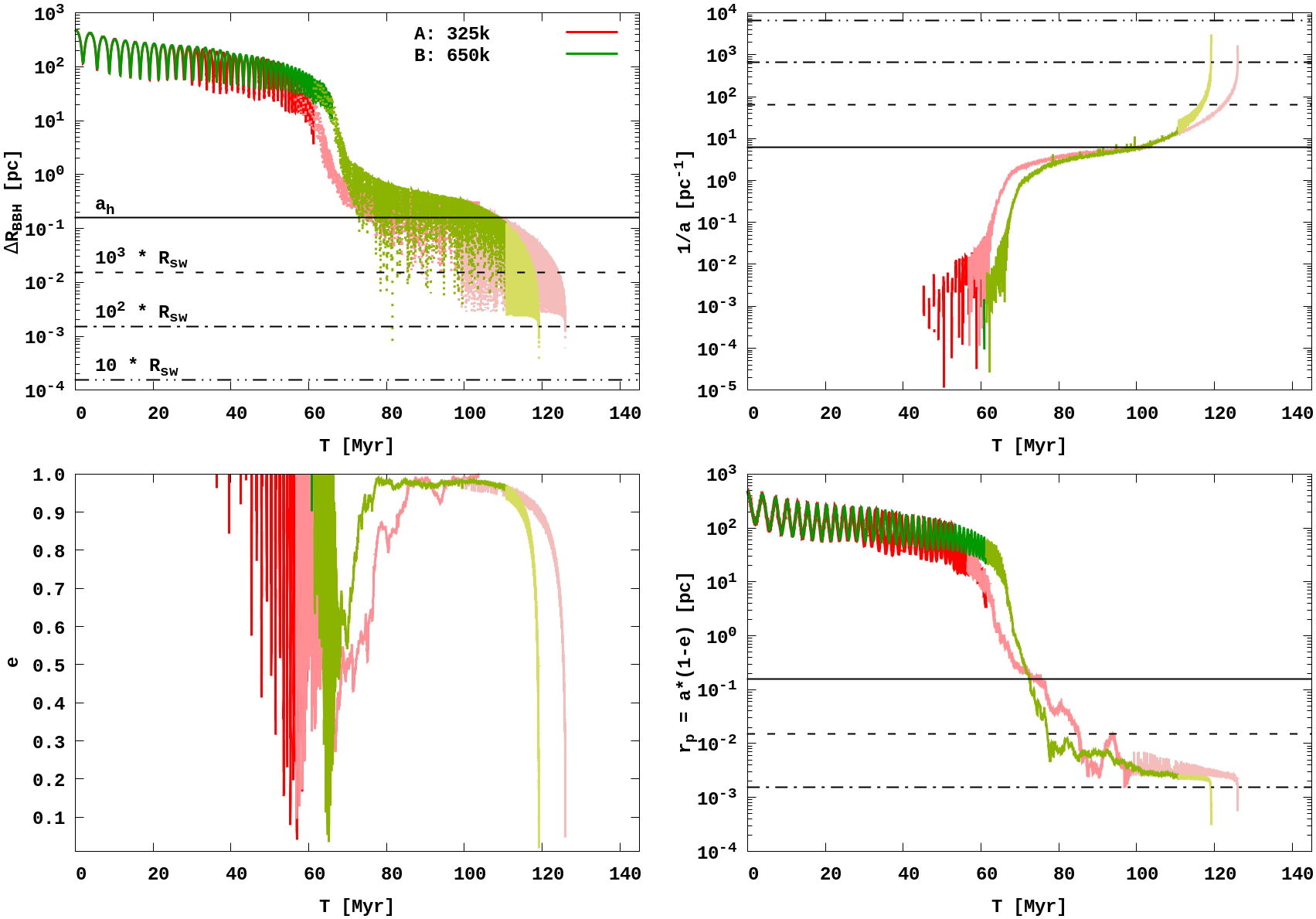}
\caption{Same as on Fig.~\ref{fig:bh-param}, but for the numerical models {\tt A$_{\tt rnd1}$} and {\tt B$_{\tt rnd1}$}. }
\label{fig:bh-param-oth}
\end{figure*}

The detailed evolution of the SMBHB parameters for our largest $N$ model simulation {\tt C$_{\tt rnd1}$}, is shown in Fig.~\ref{fig:bh-param}. The general evolution is very similar to the plots we present in Fig.~\ref{fig:7072-bound}. Here, we also show the PN phase of the dynamical evolution of SMBHB. In this case, the dissipative PN terms become significant after $\approx105$~Myr, which we can clearly see in the {\tt e} evolution, which corresponds to the SMBHB pericenter separation of $\approx1000$~R$_{\rm sw}$. 

During the final phase of the merger, the pericentre distance of SMBHB falls below $\approx10$~R$_{\rm sw}$ ({see the light blue part of the plots in Fig.~\ref{fig:bh-param}}). We stopped our runs when the separation of the SMBHB fell below $4\times({\rm R}_{\rm sw1} + {\rm R}_{\rm sw2})$. Before the very final merging stage after $\approx105$~Myr, the SMBHB completely detaches from the surrounding stellar system. Thus, for showing detailed GW frequency plots, we switched to the two-body version of the same \PGPU code with the very detailed particle output resolution, with a frequency of $\approx100$~points per orbit. The similarly detailed evolution of the other two {\tt A$_{\tt rnd1}$} and {\tt B$_{\tt rnd1}$} dynamical models are presented in Fig.~\ref{fig:bh-param-oth}. The three models have slightly different final SMBHB merging times, but all are within the range of $130\pm10$~Myr. 

During the last phase of NGC~7727 modelling, we create a more frequent output of the SMBHB parameters to enable the calculation of the system's GW signal. For the waveform calculation, we used simple GW quadrupole term expressions from \cite{Kidder1995} and also \cite{Brem2013,Sobolenko2017}: 
\begin{equation}
h^{ij} = \frac{2G\mu}{D_{\rm L}c^4}~\left[Q^{ij}+P^{0.5}Q^{ij}+PQ^{ij}+P^{1.5}Q^{ij}+~...\right].
\end{equation}
Here, $P$ is a correction term for the corresponding $\mathcal{PN}$ order, $\mu$ is the reduced mass, $D_{\rm L}$ is the luminosity distance to the galaxy, and $Q^{ij}$ is the quadrupole term. The last one can be written as
\begin{equation}
Q^{ij}=2\left[\varv^{i}\varv^{j}-\frac{GM_{\rm BH12}}{r}n^{i}n^{j}\right],
\end{equation}
where $\varv^{i}$ and $n^{i}$ are the relative velocity and normalised position vectors in this reference frame, respectively.

For illustrative purposes, we neglected the higher order terms and, in this case, we can calculate the GW tensor in the source frame simply as
\begin{equation}
h^{ij}\approx\frac{4G\mu}{D_{\rm L}c^{4}}\left[\varv^{i}\varv^{j}-\frac{GM_{\rm BH12}}{r}n^{i}n^{j}\right].
\end{equation}
We chose the virtual detector to be oriented in such a way that the coordinate axes coincide with the source frame, which allows us to not make any further coordinate transformations. We computed $h_{+}$ and $h_{\times}$ from $h^{ij}$, which give the relevant measurable strains in the `+' and `$\times$' polarisations \citep{Brem2013,Sobolenko2017}.

\begin{figure*}
\centering
\includegraphics[width=0.90\linewidth]{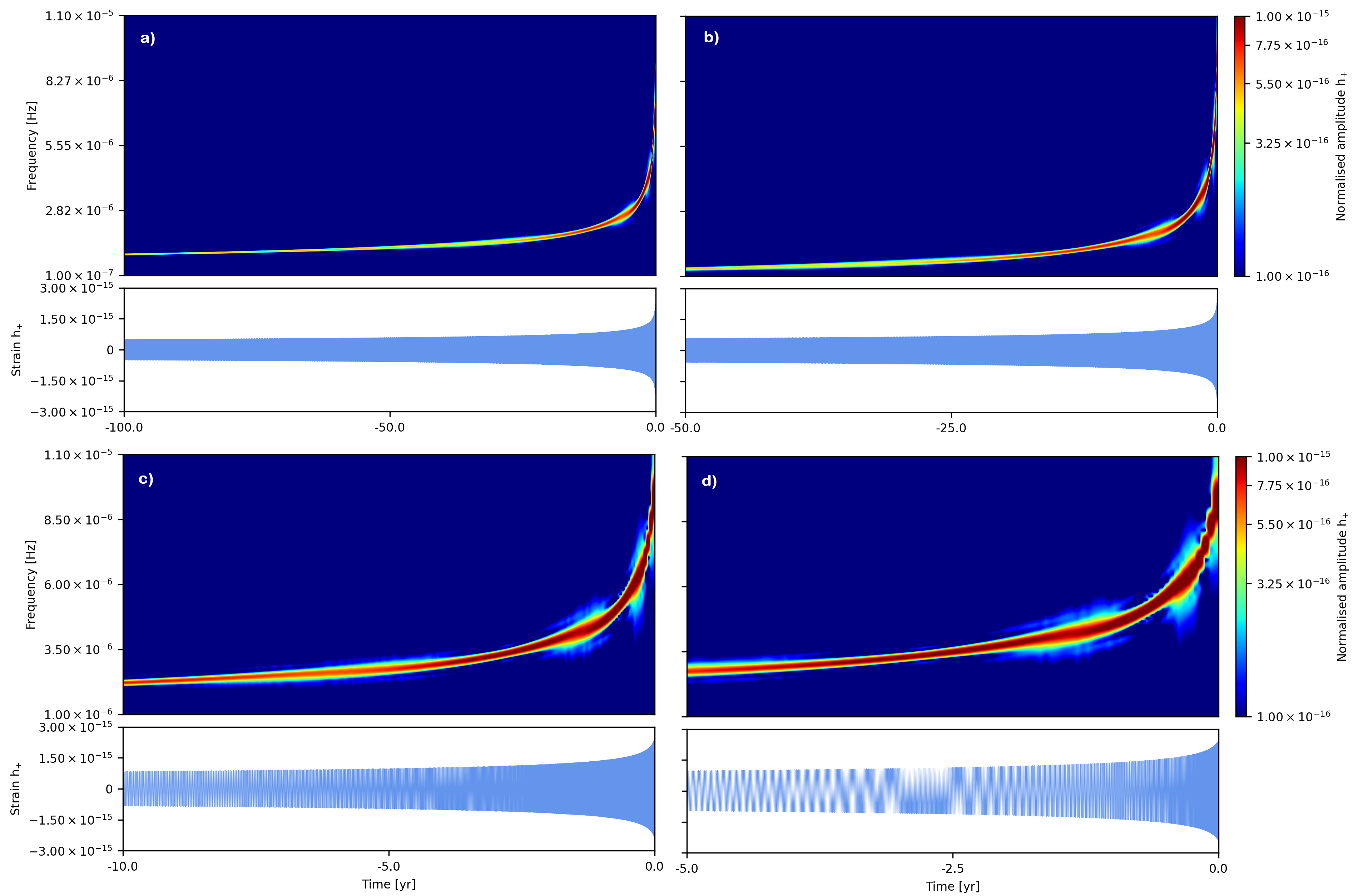}
\caption{Time-frequency representations of the strain data for predicted gravitational waveforms of $h_{+}$ polarisation from SMBHB merging of NGC~7727 for the last 100, 50, 10, and 5 yr in panels a, b, c, and d, respectively. Data refer to the {\tt C$_{\tt rnd1}$} model.} 
\label{fig:gw}
\end{figure*}

In Fig~\ref{fig:gw}, we show the waveforms and amplitude-frequency maps for the last 100, 50, 10, and 5 years for the SMBHB system before its merger. We calculated the waveforms for $h_{+}$ and $h_{\times}$ polarisations and created the amplitude-frequency maps for the final phase of our model {\tt C$_{\tt rnd1}$}. It is worth noting that the PN approximation works reasonably well only for describing the early in-spiral SMBHB. Finally, numerical relativity and perturbation theory should be used for the full waveform map of the final SMBHB merging event. 

The frequencies obtained from our simulations for the SMBHB merging events (from $\approx2\;\upmu$Hz up to 10~$\upmu$Hz, see Fig.~\ref{fig:gw}) from SMBHBs with masses $1.5\times 10^{8}\rm\;M_{\odot}$ and $6.3\times 10^{6}\rm\;M_{\odot}$ at  distances of $D_{\rm L} = 27$~Mpc are located just in between the sensitive curves of the pulsar timing array (PTA) and Laser Interferometer Space Antenna (LISA) consortiums. In other words, in principle, only during the very final phase of the merging would the GW signal from the NGC~7727 system fall within the detection range of LISA. The amplitude of the possible GW dimensionless signal strain ($\approx10^{-15} - 3\times10^{-15}$) is well above the LISA detection limit for the final 10~$\upmu$Hz frequency range. 

\subsection{The merger statistics}
We have shown that the SMBH merger would be around or just above the detection limit of the LISA interferometer. The event rates of such mergers that will be detected are clearly related to the galaxy merger rate. Early estimates of $\approx 0.008$~per~Gyr and per galaxy in the Local Universe \citep{toomre1977, tremaine1981} are supported by more recent studies based on the $\Lambda CDM$ paradigm \citep[e.g.][]{conselice2006,husko2022}.  The merger rate per galaxy increases rapidly with cosmological redshift. For example, \citet{husko2022} gives values of $\approx 10^{-2} $~per~Gyr at $ z = 0.1$, rising to $\approx 10^{-1} $ at $z = 0.5 $; whereas \citet{conselice2006} gives an estimate of $\approx 1 $~per~Gyr for a redshift range of $z \approx 1.5$ to $\le 4$. These numbers are summed over galaxy DM halos in the brackets $[10^8, 10^{12}]\rm\; M_\odot$. Based on such galactic halo statistics and JWST detection of quasar activity already at $z \approx 7$ and above,  \citet{liu2025} offers an estimate of $10^{-1} $ SMBH binary mergers per year when integrating out to $ z \simeq 5$. For systems such as NGC~7727, the timescale for the SMBH to bind and then merge $(\approx 130$~Myr) is a factor of 2 shorter than previous estimates based on dynamical friction for that system. In future works, we plan to use other galaxy configurations to explore the question of whether the merging timescale is systematically shorter than estimates based on Chandrasekhar's dynamical friction theory. We note  that identifying which late-stage mergers also host a double BH system similar to NGC~7727 is difficult observationally, as it requires resolving the sphere of influence of the black holes (see \citealt{Voggel2022} for a discussion of this issue). Generally, one may only find recent merging systems in wide-field images via their tidal features without knowing whether the merging galaxies each hosted an SMBH.  Therefore, it is not clear to what extent a significant population of candidate merging SMBH binaries might be missing in terms of the Local Universe. The new data from the DESI survey for galaxy clustering \citep{DESI_Quasars2023, DESI_Cosmo2025} should provide a solid test bed from which to further constrain the number of sources detectable by LISA. 

\section{Conclusions}\label{sec:con}

NGC~7727 is a particularly notable source because it offers a wealth of information on post-merger galaxy structures, showcasing the remnants of a smaller galaxy merging with a larger host. The presence of a secondary nucleus with an SMBH suggests complex dynamics and the survival of a stripped nuclear star cluster, a process that is often challenging to observe. 

We carried out a dynamical modelling of the evolution of the SMBH system in the dense stellar environment in the nucleus of the NGC~7727 galaxy. Our physical model was constrained based on the latest available observational data \citep{Voggel2022} for this object. We ran a set of numerical simulations using the well-tested dynamical $N$-body \PGPU code. Our code, which also includes post-Newtonian corrections, allows us to follow the dynamical evolution of the SMBHB up to a few Schwarzschild radii separations. Based on the general physical model, we constructed three numerical models with different particle number realisations, namely, 325\textit{k}, 650\textit{k}, and 1300\textit{k}, where each model also had one more additional seed of randomisations. 

At the start of the simulation, the two black holes are not gravitationally bound. The system spends more than half of its final merging time of $\approx$130~Myr in the dynamical friction phase. Then, a quickly bound binary formsm with a high eccentricity of $\approx$ 0.98, and the two-body scattering phase proceeds for  $\approx60-120$~Myr. Over the last $\approx10$~Myr, the BHs separation is seen to be rapidly shrinking due to the gravitational wave emission. The final merging time in our model is around $130\pm10$~Myr when starting from the actual separation today. Our study shows the principal possibility of the NGC~7727 binary black holes merging on a relatively short timescale. Such sources of gravitational waves are potential targets for the future LISA campaign during the final merging in the 10~$\upmu$Hz frequency range.

\begin{acknowledgements}

{The authors thank the anonymous referee for a very constructive report and suggestions that helped significantly improve the quality of the manuscript.}

The authors gratefully acknowledge computing resources used for this modelling on the Gauss Center for Supercomputing e.V. (GCS), through the John von Neumann Institute for Computing (NIC), with the GCS supercomputers JUWELS-booster at Julich Supercomputing Centre (JSC), Germany.

A significant part of the simulations was also performed on the Luxembourg national supercomputer MeluXina (project p200574). The authors appreciate the LuxProvide teams for their expert support.

PB, MI, OV, and MS thank the support from the special program of the Polish Academy of Sciences and the U.S. National Academy of Sciences under the Long-term program to support Ukrainian research teams, grant No.~PAN.BFB.S.BWZ.329.022.2023.

PB acknowledges the support of the National Science Foundation of China (NSFC) under grant No.~12473017. 

We thank the ``Development of the concept for the first Kazakhstani orbital cislunar telescope - Phase I'', financed by the Ministry of Science and Higher Education of the Republic of Kazakhstan, No.~BR24992759.

This collaboration was made possible thanks to the support from the French-Ukrainian ``DNIPRO'' through grant number Ts~17/2--1 awarded in 2022-23 jointly to PB, MI, and CMB: we are grateful to all these agencies. 

\end{acknowledgements}

\bibliographystyle{mnras}  
\bibliography{ngc-7727}   



\label{LastPage}
\end{document}